\documentclass[twoside,11pt]{article}

\usepackage[accepted]{melba}

\usepackage{amsmath,amsfonts}

\usepackage{soul}
\newcommand{\singlequotes}[1]{\lq#1\rq}
\DeclareMathOperator*{\argmax}{arg\,max}
\DeclareMathOperator*{\argmin}{arg\,min}
\usepackage{array}

\melbaid{2024:002}  
\doi{https://doi.org/10.59275/j.melba.2024-geb5}
\melbaauthors{Epstein et al.}  
\volume{2}
\firstpageno{586}  
\melbayear{2024}  
\datesubmitted{06/2023}  
\datepublished{02/2024}  

\ShortHeadings{Training labels for qMRI parameter estimation}{Epstein et al.}

\title{Choice of training label matters: how to best use \\ deep learning for quantitative MRI parameter estimation}

\author{\firstname Sean C. \surname Epstein \email rmapcep@ucl.ac.uk \\  
	\addr Centre for Medical Image Computing, University College London, London, UK
	\AND
	\name Timothy J. P. Bray \email t.bray@ucl.ac.uk \\
	\addr Centre for Medical Imaging, University College London, London, UK
    \AND
    \name Margaret Hall-Craggs \email m.hall-craggs@ucl.ac.uk \\
	\addr Centre for Medical Imaging , University College London, London, UK
    \AND
    \name Hui Zhang \email gary.zhang@ucl.ac.uk \\
	\addr Centre for Medical Image Computing, University College London, London, UK
}

\begin{document}

\maketitle

\begin{abstract}
	Deep learning (DL) is gaining popularity as a parameter estimation method for quantitative MRI. A range of competing implementations have been proposed, relying on either supervised or self-supervised learning. Self-supervised approaches, sometimes referred to as unsupervised, have been loosely based on auto-encoders, whereas supervised methods have, to date, been trained on groundtruth labels. These two learning paradigms have been shown to have distinct strengths. Notably, self-supervised approaches offer lower-bias parameter estimates than their supervised alternatives. This result is counterintuitive – incorporating prior knowledge with supervised labels should, in theory, lead to improved accuracy. In this work, we show that this apparent limitation of supervised approaches stems from the naïve choice of groundtruth training labels. By using \textit{intentionally-non-groundtruth training labels}, pre-computed via independent maximum likelihood estimation, we show that the low-bias parameter estimation previously associated with self-supervised methods can be replicated – and improved on – within a supervised learning framework. This approach sets the stage for a single, unifying, deep learning parameter estimation framework, based on supervised learning, where trade-offs between bias and variance are made by careful adjustment of training label.
	Our code is available at~\url{https://github.com/seancepstein/training_labels}.
\end{abstract}

\begin{keywords}
	quantitative MRI, diffusion MRI, deep learning
\end{keywords}
\section{Introduction}
Magnetic resonance imaging (MRI) is widely regarded as the premier clinical imaging modality, in large part due to the unparalleled range of contrast mechanisms available to it. Conventional MRI exploits this contrast in a purely qualitative manner: images provide only \textit{relative} information, such that voxel intensities are only meaningful in the context of their neighbours. In contrast, quantitative MRI (qMRI) provides quantitative images, where voxel intensities are directly, and meaningfully, related to underlying tissue properties. Compared to conventional MRI, this approach promises increased reproducibility, interpretability, and tissue insight, at the cost of time-intensive image acquisition and post-processing \citep{Cercignani2018}.

One of the biggest time and resource bottlenecks in post-processing is parameter estimation, whereby a signal model is fit to the intensity variation across \textit{multiple} MR images acquired at different experimental settings. Each voxel requires its own independent model fit: solving for the signal model parameters that best described the single voxel's data. The computational cost of this curve-fitting process, which scales with both voxel number and model complexity, has become a bottleneck for modern qMRI experiments.

Accelerating curve fittings with deep learning (DL) was first proposed more than 30 years ago \citep{Bishop1992}, but has only recently gained popularity within the qMRI community \citep{Golkov2016a, Bertleff2017, Liu2020,Barbieri2020,Palombo2020}. Just like traditional methods, DL relies on model fitting, but the model being fit is a fundamentally different one. Instead of fitting a qMRI \textit{signal model} to a single voxel of interest (i.e. curve fitting), DL methods fit (``train") a \textit{deep neural network (DNN) model} to an ensemble of training voxels. This model maps a single voxel's signal to its corresponding qMRI parameters; the unknowns in its fitting are network weights, rather than qMRI parameters. Once this DNN model has been fit to (``trained on") the training data, parameter estimation is reduced to simply applying it to new unseen data, one voxel at a time. This approach offers two broad advantages over traditional fitting: (1) computational cost is amortised: despite being more computationally expensive than one-voxel signal model fitting, DL training only needs to be performed once, for any number of voxels; once trained, networks provide near-instantaneous parameter estimates on new data, and (2) computational cost is front-loaded: model training can be performed away from the clinic, before patient data is acquired.

To date, most DL qMRI fitting methods have been implemented within a supervised learning framework \citep{Golkov2016a,Bertleff2017,Yoon2018,Liu2020,Palombo2020,Aliotta2021,Yu2021,Gyori2022}. This approach trains DNNs to predict groundtruth qMRI model parameters from noisy qMRI signals. When compared to conventional fitting, this approach has been found to produce high bias, low variance parameter estimates \citep{Grussu2021,Gyori2022}.

An alternative class of DL methods has also been proposed, sometimes referred to as unsupervised learning \citep{Barbieri2020,Kaandorp2021}, but more accurately described as self-supervised \citep{pmlMurphy}. In this framework, training labels are not explicitly provided, but are instead extracted by the network from its training input. This label generation is designed such that the network learns to predict signal model parameters corresponding to noise-free signals that most-closely approximate noisy inputs. This self-supervised approach has been found to produce similar results to conventional non-DL fitting, i.e. lower bias and higher variance than its groundtruth-labelled supervised alternative \citep{Barbieri2020,Grussu2021}.

From an information theoretic standpoint, the comparison between supervised and self-supervised performance raises an obvious unanswered question. How can it be that supervised methods, which provide strictly more information during training than their self-supervised counterparts, produce more biased parameter estimates?

In this work we answer this question by showing that this apparent limitation of supervised approaches stems purely from the selection of groundtruth training labels. By using \textit{intentionally-non-groundtruth training labels}, pre-computed via independent maximum likelihood estimation, we show that the low-bias parameter estimation previously associated with self-supervised methods can be replicated – and improved on – within a supervised learning framework.

This approach sets the stage for a single, unifying, deep learning parameter estimation framework, based on supervised learning, where trade-offs between bias and variance can be made, on an application-specific basis, by careful adjustment of training label.

The rest of the paper is organized as follows: Section~\ref{sec:theory} describes existing DL parameter estimation approaches, our proposed method, and how they relate to each other; Section~\ref{sec:method} describes the evaluation of our method and its comparison to the state of the art; Section~\ref{sec:results} contains our findings; and Section~\ref{sec:conclusions} summarizes the contribution and discusses future work.

\section{Theory} \label{sec:theory}
Quantitative MRI extracts biomarkers $y$ from MR data $x$, producing quantitative spatial maps. We here describe existing voxelwise approaches to this problem (conventional fitting and DL alternatives) as well as our proposed novel method.

\subsection{Conventional iterative fitting}

This method, which relies on maximum likelihood estimation (MLE), extracts biomarkers by performing a voxelwise model fit every time new data is acquired. An appropriate signal model $M$ is required, parameterised by $n_y$ parameters of interest; for each combination of $y$, the probability of observing the acquired data $x$ is known as the likelihood $L$ of those parameters:

\begin{equation}
    L(x,z|y,\epsilon) = \prod_{i=1}^{n_z} P(x_i,z_i|y,\epsilon)
\end{equation}

\noindent for $n_z$ acquisitions from sampling scheme $z$ and noise model $\epsilon$. The model parameters $\hat{y}$ which maximise the likelihood $L$ are assumed to best represent the tissue contained within the voxel of interest:

\begin{equation}
    \hat{y} = \argmax_y L(x,z|y,\epsilon)
\end{equation}

\noindent Under a Gaussian noise model, this likelihood maximization reduces to the commonly-used non-linear least squares (NLLS):

\begin{equation}
    \hat{y} = \argmin_y \sum_{i=1}^{n_z} \lVert M(z_i|y) - x_i \rVert^2
\end{equation}

\noindent under the assumption of signal model $M$ associated with groundtruth biomarkers $y_{gt}$, sampling scheme $z$, and noise $\epsilon$:

\begin{equation}
    x = M(z|y_{gt}) + \epsilon
\end{equation}

\noindent Each of these optimisations has $n_y$ unknowns, which are solved for independently across different voxels; the computational cost scales linearly with the number of voxels $n_v$.

Developments in qMRI acquisition and analysis have led to increased (i) image spatial resolution (i.e. greater $n_v$) and (ii) model complexity (i.e. greater $n_y$), such that conventional MLE fitting has become increasingly computationally expensive.

\subsection{Existing deep learning methods}
\label{sec:existing}

Deep learning approaches address this by reframing $n_v$ independent problems into a \textit{single} global model fit: learning the function $\mathcal{F}$ that maps any $x$ to its corresponding $y_{gt}$: 

\begin{equation}
    y_{gt} = \mathcal{F}(x)
\end{equation}

Deep neural networks aim to approximate this function by composing a large but finite number of building-block functions, parametrised by $n_p$ network parameters $p$ (``weights"):

\begin{equation}
    \hat{y} = \hat{\mathcal{F}}(x|\hat{p})
\end{equation}

In this context, model fitting (``training"), is performed over network parameters $p$ and involves maximising $\hat{\mathcal{F}}$'s mean performance over a large set of training examples; the trained network is defined by the best-fit parameters $\hat{p}$. This fitting problem, whilst more computationally expensive to solve than any individual voxel ($n_v=1$) MLE, is only tackled once; once $\hat{\mathcal{F}}$ is learnt, it can be applied at negligible cost to new, unseen, data. This promise of rapid, zero-cost parameter estimation has led to the development of two broad classes of DL-based parameter estimation methods.

\textit{Supervised\textsubscript{GT}} methods approximate $\mathcal{F}$ by minimising the difference between a large number of noise-free training labels (groundtruth parameter values) and corresponding network outputs (noise-free parameter estimates); training loss is calculated in the parameter space $Y$:

\begin{equation}
    \text{{Supervised\textsubscript{GT} training loss}} = \sum_{i=1}^{n_{train}} \lVert W \cdot (\hat{y}_{i} - y_{gt,i})\rVert^2
\end{equation}

\noindent where $n_{train}$ is the number of training samples and $W$ is a tunable weight matrix which accounts for magnitude differences in signal model parameters. $W$ is generally a diagonal matrix, with each diagonal element $W_{ii}$ corresponding to the relative weighting of qMRI parameter $y_i$; setting $W$ as the identity matrix equally weights all parameters in the training loss.

These methods produce higher bias, lower variance parameter estimation than conventional MLE fitting \citep{Grussu2021,Gyori2022} and, by adjusting $W$, can be tailored to selectively boost estimation performance on a subset of the parameter space $Y$. 

In contrast, \textit{Self-supervised} methods compute training loss within the signal space $X$, by minimising the difference between network inputs (noisy signals) and a filtered representation of network outputs (noise-free signal estimates): 

\begin{equation} \label{eq:selfsupervisedloss}
    \text{{Self-supervised training loss}} = \sum_{i=1}^{n_{train}} \lVert M(z|\hat{y}_{i}) - x_{i} \rVert^2
\end{equation}

\noindent These methods, which perform similarly to conventional MLE fitting, produce lower bias, higher variance parameter estimation than \textit{Supervised\textsubscript{GT}}\citep{Grussu2021,Barbieri2020}. Unlike \textit{Supervised\textsubscript{GT}}, the relative loss weighting of different signal model parameters is dictated by sampling scheme $z$.

Under Gaussian noise conditions, single-voxel \textit{Self-supervised} loss (i.e. minimising the sum of squared differences between a noisy signal and its noise-free signal estimate) is indistinguishable from the corresponding objective function in conventional fitting.

In contrast, under the Rician noise conditions encountered in MRI acquisition\citep{Gudbjartsson1995}, \textit{Self-supervised} training loss no longer matches conventional fitting. Indeed, the sum of squared errors between noisy signals and noise-free estimates is not an accurate difference metric in the presence of Rician noise.

To summarise: existing supervised DL techniques are associated by high estimation bias, low variance, and end-user flexibility; in contrast, self-supervised methods have lower bias, higher variance, but are limited by the fact their loss is calculated in the signal space $X$.

\subsection{Proposed deep learning method}
\label{sec:proposed}

\begin{figure}[!h]
\includegraphics[width=\textwidth]{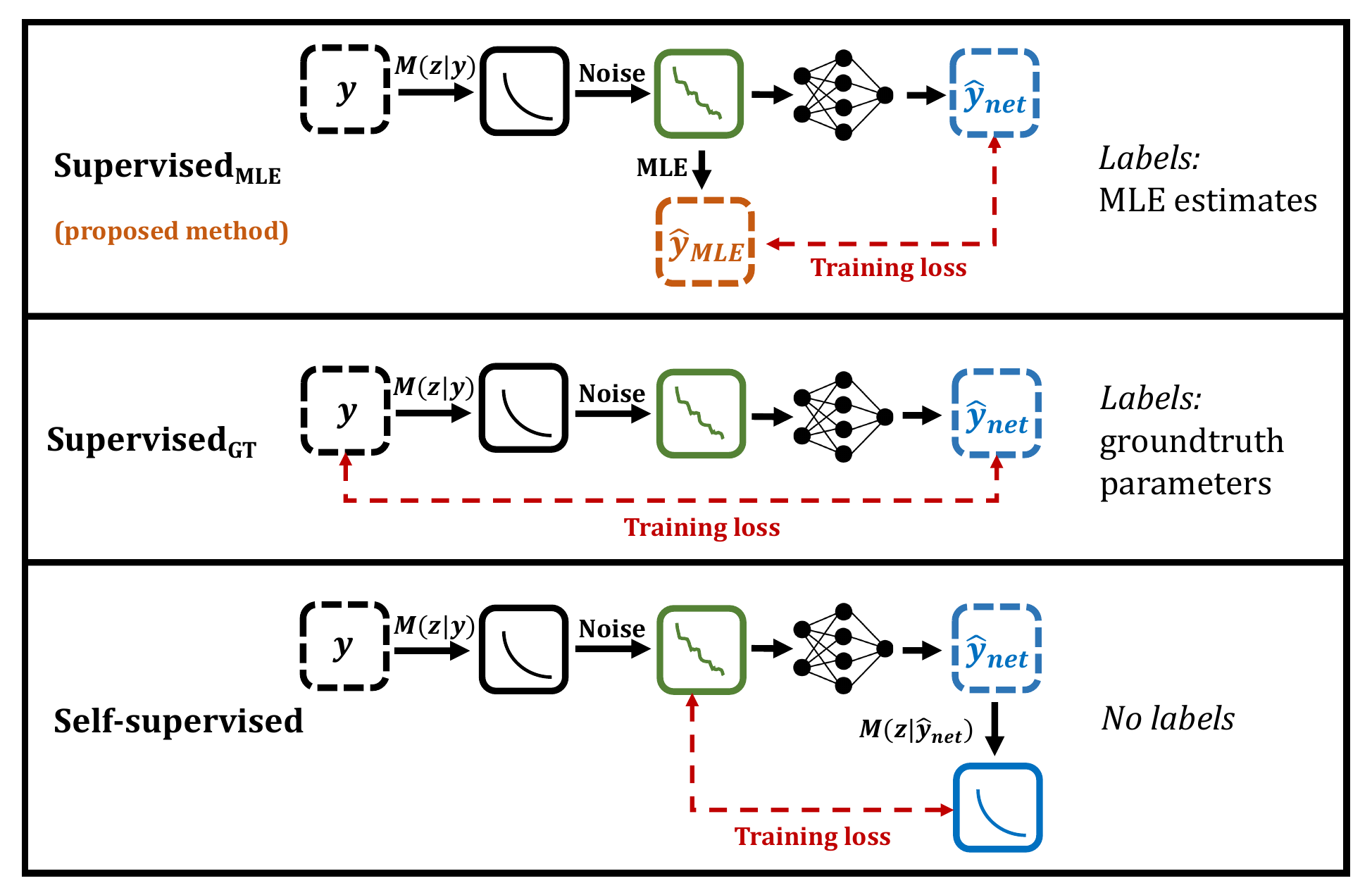}
\caption{Comparison between our proposed method (\textit{Supervised\textsubscript{MLE}}) and existing supervised and self-supervised approaches.}
\label{fig:methodcomparison}
\end{figure}

In light of this, we propose \textit{Supervised\textsubscript{MLE}}, a novel parameter estimation method which combines the advantages of \textit{Supervised\textsubscript{GT}} and \textit{Self-supervised} methods. This method is contrasted to existing techniques in Fig \ref{fig:methodcomparison}.

This method mimics \textit{Self-supervised}'s low-bias performance by learning a regularised form of conventional MLE, but does so in the parameter space $Y$, within a supervised learning framework. This addresses the limitations of \textit{Self-supervised}: Rician noise modelling is incorporated, and parameter loss weighting is not limited by sampling scheme $z$.

Our method learns $\hat{\mathcal{F}}$ by training on noisy signals paired with conventional MLE labels. These labels act as proxies for the groundtruth parameters we wish to estimate: 

\begin{equation}
    \text{{Supervised\textsubscript{MLE} training loss} } = \sum_{i=1}^{n_{train}} \lVert W \cdot (\hat{y}_{i} - y_{MLE,i})\rVert^2
\end{equation}

where $y_{MLE,i}$ is the maximum likelihood estimate associated with the $i^{th}$ training sample.

Our method offers one final advantage over \textit{Self-supervised} approaches. In addition to the parameter estimation improvements relating to noise model correction and parameter loss weighting, it naturally interfaces with \textit{Supervised\textsubscript{GT}}. In so doing, it presents the opportunity to combine low-bias and low-variance methods into a single, tunable hybrid approach, by a simple weighted sum of each method's loss function:

\begin{equation}
    \text{{Hybrid training loss} } = \alpha \cdot \text{{Supervised\textsubscript{MLE} loss} } + (1-\alpha) \cdot \text{{Supervised\textsubscript{GT} loss}}
\end{equation}

\section{Experimental evaluation}
\label{sec:method}

Three classes of network were investigated and compared: \textit{Supervised\textsubscript{GT}}, \textit{Self-supervised}, and \textit{Supervised\textsubscript{MLE}}, as described in Fig \ref{fig:methodcomparison}. Additionally, to control for differences in loss function weighting between supervised and unsupervised methods, \textit{Self-supervised} was converted into supervised form by training \textit{Supervised\textsubscript{MLE}} on Gaussian-model based MLE labels. All models are summarised in Table \ref{table:networks}.

\begin{table}[!h]
\centering
\caption{Summary of evaluated parameter estimation networks. $Y$ denotes parameter space; $X$ denotes signal space.}
\begin{tabular}{|m{10em} | m{7em} | m{6em} | m{10em} |}
 \hline
 \textbf{Network name} & \textbf{Loss space} & \textbf{Label} & \textbf{Label noise model} \\
 \hline
 \textit{Supervised\textsubscript{GT}} & $Y$  & Groundtruth & N/A \\
\hline
\textit{Self-supervised} & $X$  & N/A & N/A \\
\hline
\textit{Supervised\textsubscript{MLE, Rician}} & $Y$  & MLE & Rician \\
\hline
\textit{Supervised\textsubscript{MLE, Gaussian}} & $Y$  & MLE & Gaussian \\
\hline
\end{tabular}
\label{table:networks}
\end{table}

All networks were trained and tested on the same datasets; differences in performance can be attributed solely to differences in loss function formulation and training label selection. 

\subsection{Signal model}

The intravoxel incoherent motion (IVIM) model \citep{LeBihan1986} was investigated as an exemplar 4-parameter non-linear qMRI model which poses a non-trivial model fitting problem and is well-represented in the DL qMRI literature \citep{Bertleff2017,Barbieri2020,Kaandorp2021,Mastropietro2022,Rozowski2022}:

\begin{equation}
    S(b|S_0,f,D_{slow},D_{fast}) = S_0 (f e^{-b (D_{fast} + D_{slow})} + (1-f) e^{-b D_{slow}})
    \label{eq:ivim}
\end{equation}

\noindent where $S$ corresponds to the signal model $M$, $b$ corresponds to the sampling scheme $z$, and $[S_0, f, D_{slow}, D_{fast}]$ corresponds to the parameter-vector $y$. In physical terms, IVIM is a two-compartment diffusion model, wherein signal decay arises from both molecular self-diffusion (described by $D_{slow}$) and perfusion-induced \singlequotes{pseudo-diffusion} (described by $D_{fast}$). In Equation \ref{eq:ivim}, $S_0$ is an intensity normalisation factor and $f$ denotes  the signal fraction corresponding to the perfusing compartment.

\subsection{Network architecture}

Network architecture was harmonised across all network variants, and represents a common choice in the existing qMRI literature \citep{Barbieri2020}: 3 fully connected hidden layers, each with a number of nodes matching the number of signal samples $z$ (i.e. b-values), and an output layer with a number of nodes matching the number of model parameters. Wider (150 nodes per layer) and deeper (10 hidden layers) networks were investigated and found to have equivalent performance, during both training and testing, at the cost of increased training time. All networks were implemented in Pytorch 1.9.0 with  exponential  linear  unit  activation  functions \citep{Clevert2015}; ELU performance is similar to ReLU, but is more robust to poor network weight initialisation.

\subsection{Training data} \label{sec:training-data}

Training datasets were generated at \textit{SNR} $= [15,30]$ to investigate parameter estimation performance at both high and low noise levels. At each SNR, 100,000 noise-free signals were generated from uniform IVIM parameter distributions ($S_0 \in [0.8, 1.2]$, $f \in [0.1, 0.5]$, $D_{slow} \in [0.4, 3.0] 10^{-3} mm^2 /s$, $D_{fast} \in [10, 150] 10^{-3} mm^2 /s$, representing realistic tissue values), sampling them with a real-world acquisition protocol \citep{Zhao2015} ($b = [0, 10, 20, 30, 50, 80, 100, 200, 400, 800]$ $s / mm^2$), and adding Rician noise. Training data generative parameters were drawn from uniform, rather than in-vivo, parameter distributions to minimise bias in network parameter estimation\citep{Gyori2022}. Data were split 80/20 between training and validation. MLE labels were calculated using a bound-constrained non-linear fitting algorithm, implemented with \textit{scipy.optimize.minimize}, using either Rician log-likelihood (for \textit{Supervised\textsubscript{MLE, Rician}}) or sum of squared errors (for \textit{Supervised\textsubscript{MLE, Gaussian}}) as fitting objective function. This algorithm was initialised with groundtruth values (i.e. generative $y$) to improve fitting robustness and avoid local minima. Training/validation samples associated with \singlequotes{poor} MLE labels (defined as lying on the boundary of the bound-constrained estimation space) were held out during training and ignored during validation.

\subsection{Network training}

Network training was performed using an Adam optimizer (learning rate = 0.001, betas = (0.9, 0.999), weight decay=0) as follows: \textit{Supervised\textsubscript{GT}} (at SNR 30) was trained 16 times on the same data, each time initialising with different network weights, to improve robustness to local minima during training. From this set of trained networks, a single \textit{Supervised\textsubscript{GT}} network was selected on the basis of validation loss. The trained weights of this selected network were subsequently used to initialise all other networks; in this way, any differences in network performance could be solely attributed to differences in training label selection and training loss formulation. In the case of supervised loss formulations, the inter-parameter weight vector $W$ was chosen as the inverse of each parameter's mean value over the training set, to obtain equal loss weighting across all four IVIM parameters.

\subsection{Testing data} \label{groundtruths}

Networks were tested on both synthetic and real qMRI data. The synthetic approach offers (i) known parameter groundtruths to assess estimation against, (ii) arbitrarily large datasets, and (iii) tunable data distributions, but is based on possibly simplified qMRI signals. This approach was used to assess parameter estimation performance in a controlled, rigorous manner; real data was subsequently used to validate the trends observed in silico.

Synthetic data was generated with sampling, parameter distributions, and noise levels matching those used in network training. The IVIM parameter space in which the networks were trained was uniformly sub-divided 10 times in each dimension, to analyse estimation performance as a function of parameter value. At each point in the parameter space, 500 corresponding noisy signals were generated and used to test network performance, accounting for variation under noise repetition.

Real data was acquired from the pelvis of a healthy volunteer, who gave informed consent, on a wide-bore 3.0T clinical system (Ingenia, Philips, Amsterdam, Netherlands), 5 slices, 224 x 224 matrix, voxel size = 1.56 x 1.56 x 7mm, TE = 76ms, TR = 516ms, scan time = 44s per 10 b-values listed in \autoref{sec:training-data}. For the purposes of assessing parameter estimation methods, we obtained gold standard voxelwise IVIM parameter estimates from a supersampled dataset (16-fold repetition of the above acquisition, within a single scanning session, generating 160 b-values, total scan time = 11m44s). Conventional MLE was performed on this supersampled data to produce best-guess ``groundtruth"  parameters. During testing, the supersampled dataset was split into 16 distinct 10 b-value acquisitions, each corresponding to a single realistic clinical acquisition. All images were visually confirmed to be free from motion artefacts.
The mismatch in parameter distributions between this in-vivo data (highly non-uniform) and the previously-described synthetic data (uniform by construction) limited the scope for validating our in-silico results. To address this, a final synthetic testing dataset was generated from the in-vivo MLE-derived ``groundtruth" parameters, and was used for direct comparison between real and simulated data.

\subsection{Evaluation metrics}

Parameter estimation performance was evaluated using 3 key metrics: (i) mean bias with respect to groundtruth, (ii) mean standard deviation under noise repetition, and (iii) root mean squared error (RMSE) with respect to groundtruth. RMSE is the most commonly used metric to evaluate estimation performance \citep{Barbieri2020,Bertleff2017}, but is limited in its ability to disentangle accuracy and precision; to this end, mean bias and standard deviation were used as more specific measures of network performance.

It is important to note that \textit{all} methods were assessed with respect to groundtruth qMRI parameters, \textit{even those trained on MLE labels}. For these methods, the training and validation loss (MLE-based) differed from the reported testing loss (groundtruth-based).

\section{Results \& discussion}
\label{sec:results}

This section summarises our main findings and discusses the advantages offered by the parameter estimation method we propose.

\subsection{Comparison of parameter estimation methods} \label{summary}

\begin{figure}[!h]
  \includegraphics[width=\textwidth]{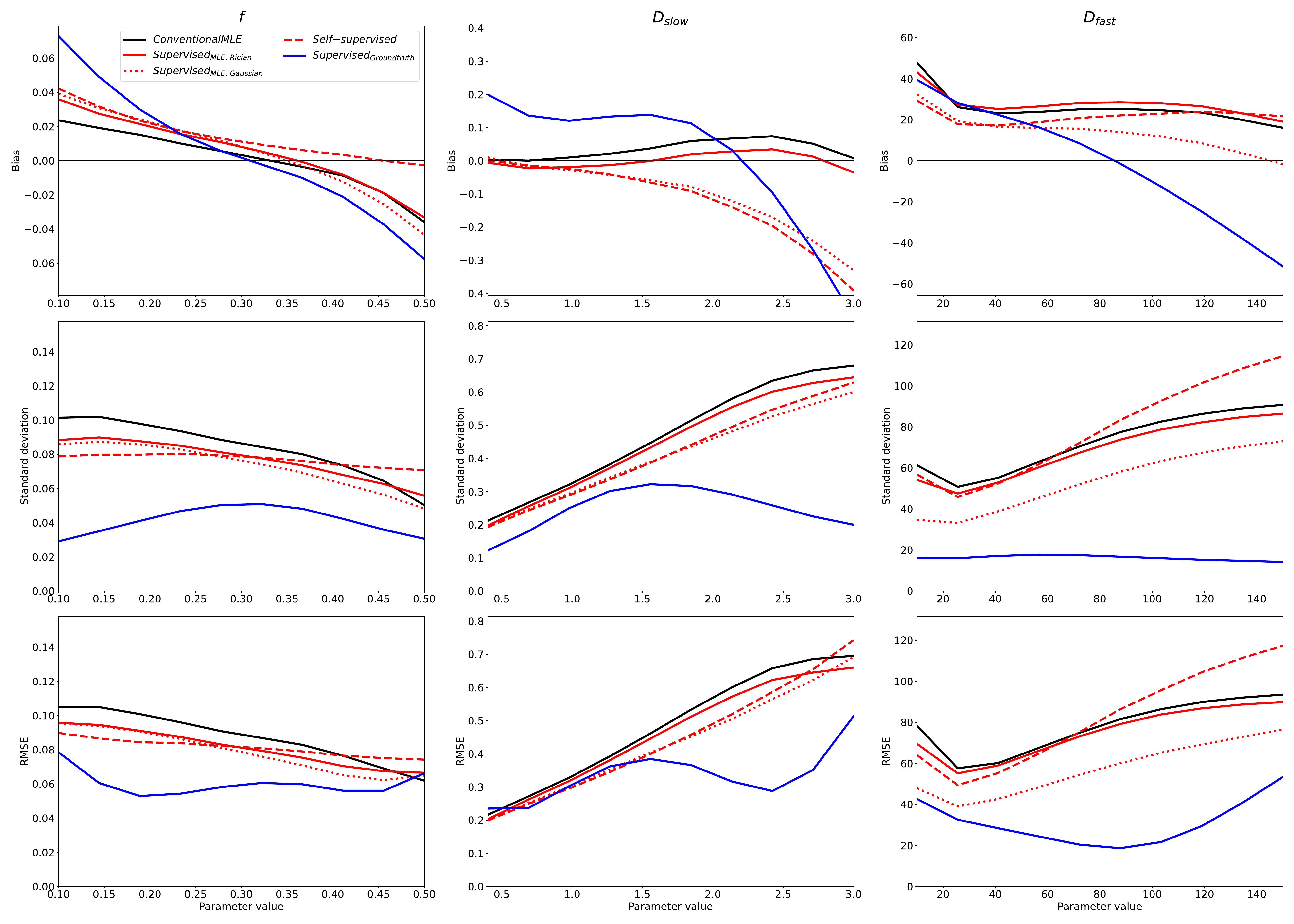}
  \caption{Parameter estimation performance at low SNR (15) as a function of groundtruth parameter $Y$. Performance summarised by bias \& RMSE with respect to groundtruth and standard deviation with respect to noise repetition. Conventional MLE fitting is provided as a non-DNN reference standard. For the sake of visualisation, each plotted point represents marginalisation over all non-specified ${Y}$ dimensions.}
  \label{fig:lowsnr}
\end{figure}

\begin{figure}[!h]
  \includegraphics[width=\textwidth]{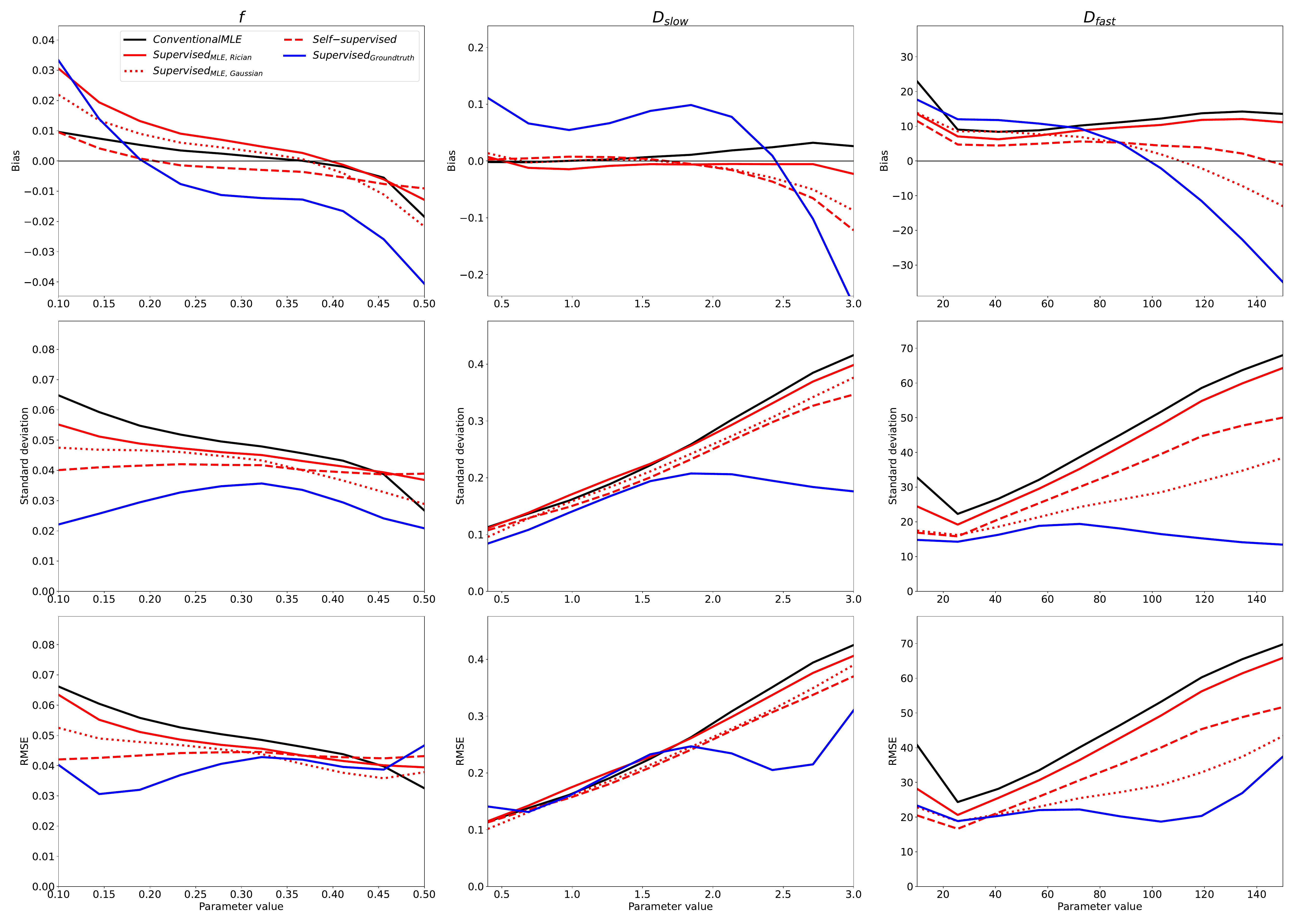}
  \caption{Parameter estimation performance, visualised as in Figure \ref{fig:lowsnr}, but for high SNR (30) data.}
  \label{fig:highsnr}
\end{figure}

The relative performance of all previously-discussed parameter estimation methods is summarised in Figures \ref{fig:lowsnr} and \ref{fig:highsnr}. These figures show the bias, variance (represented by its square root: standard deviation), and RMSE of parameter estimates with respect to groundtruth values, reported for each model parameter as a function of its value over the synthetic test dataset; each plotted point represents an average over 500 noise instantiations and a marginalisation over all non-visualised parameters. Marginalisation was required for visualisation of a 4-dimensional parameter space, but was confirmed to be representative of the entire, non-marginalised space, as discussed in §\ref{marginalisation}.

In keeping with previously reported results, we show a bias/variance trade-off between different parameter estimation methods. Conventional MLE fitting is provided as a reference (plotted in black). Approaches which, on a theoretical level, approximate conventional MLE (\textit{Self-supervised} and \textit{Supervised\textsubscript{MLE}}, plotted in red), are generally associated with low bias, high variance, and high RMSE, whereas groundtruth-labelled supervised methods (plotted in blue) exhibit lower variance and RMSE at the cost of increased bias. 

\begin{figure}[!h]
\centering
  \includegraphics[width=\textwidth]{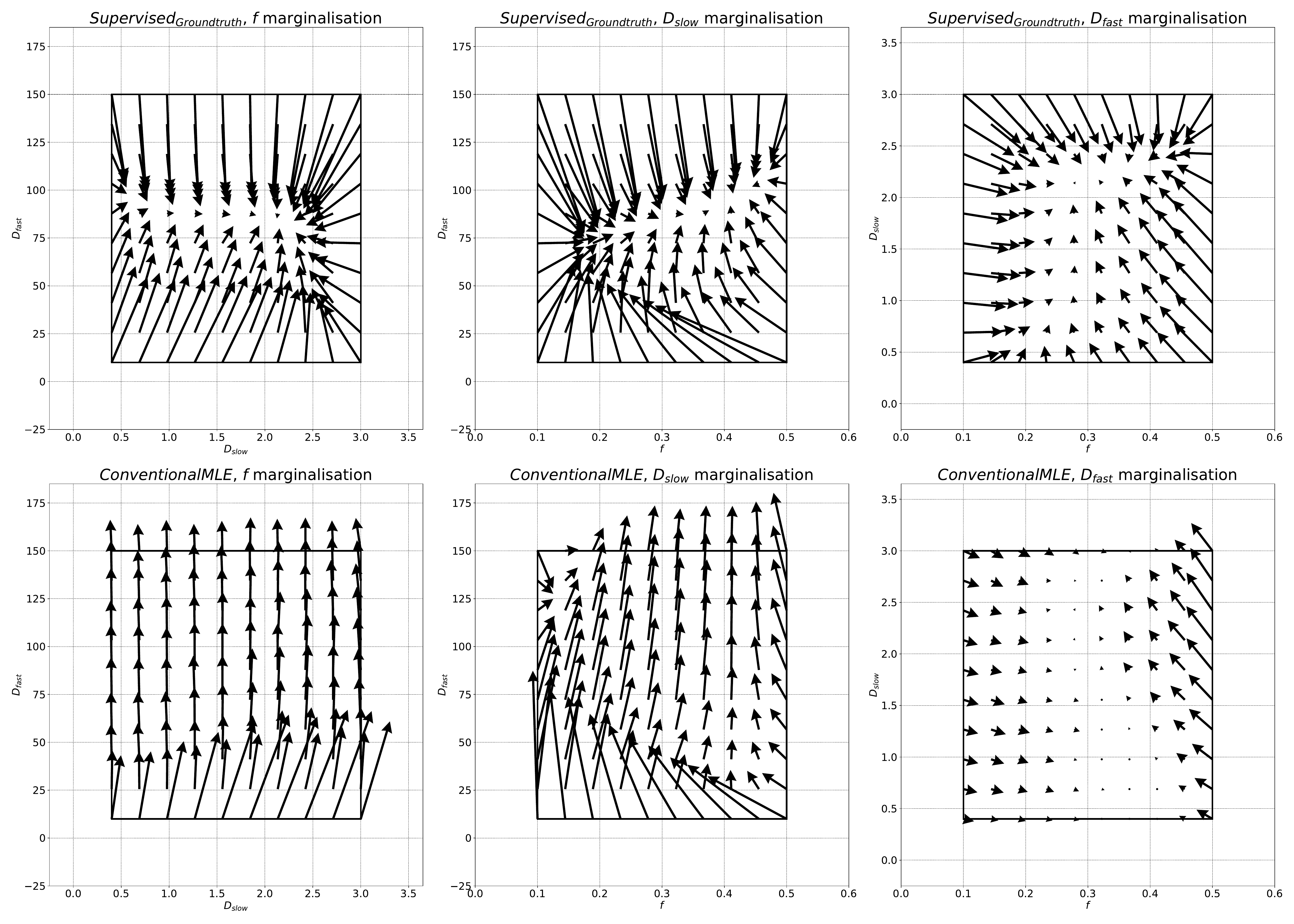}
  \caption{Comparison between \textit{Supervised\textsubscript{GT}} and reference conventional MLE fitting, expressed in terms of estimation bias and information compression at low SNR (15). Arrows represent the mean mapping from $Y$ to $\hat{Y}$, averaged over noise, as a function of parameter space $Y$. For the sake of visualisation, each plotted point represents marginalisation over all non-specified ${Y}$ dimensions.}
  \label{fig:quiver1}
\end{figure}

Increases in bias, if consistent across parameter space $Y$, do not necessarily reduce sensitivity to differences in underlying tissue properties. However, we show that $Supervised_{GT}$ is associated with bias that \textit{varies significantly} as a function of groundtruth parameter values. This results in a reduction in information content, visualised as the \textit{gradient} of the bias plots (top row) in Fig \ref{fig:lowsnr}. The more negative the gradient, the more parameter estimates are concentrated in the centre of the parameter estimation space $\hat{X}$, and the lower the ability of the method to \textit{distinguish} differences in tissue properties. This information loss can be seen in Fig \ref{fig:quiver1}, which compares $Supervised_{GT}$ to conventional MLE fitting, and shows the compression in $\hat{X}$ over the groundtruth parameter-space $X$.

\subsection{Validation against clinical data}

\begin{figure}[!h]
  \includegraphics[width=\textwidth]{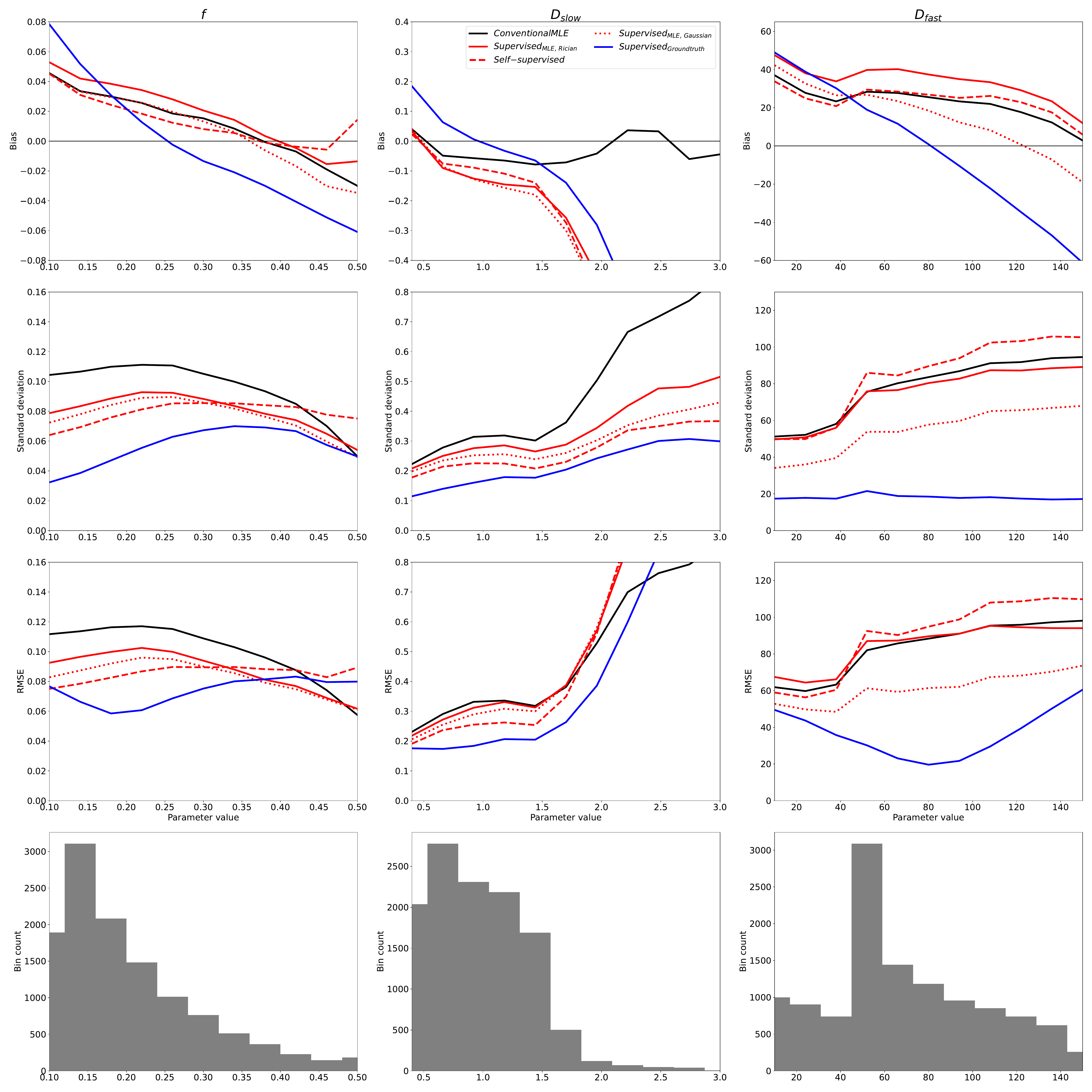}
  \caption{In vivo parameter estimation performance of networks trained on low SNR (15) synthetic data, as a function of supersampling-derived reference parameter values. The first three rows summarise performance by showing bias \& RMSE with respect to reference value and standard deviation with respect to noise repetition, marginalised over all non-specified ${Y}$ dimensions. The bottom row shows the distribution of reference parameter values across the parameter range being visualised.}
  \label{fig:clinicallowsnr}
\end{figure}

The above trends, found in simulation, were also observed in real-world data. Fig \ref{fig:clinicallowsnr} shows the bias, variance,
and RMSE of parameter estimates with respect to ``groundtruth" values (obtained from the supersampled dataset described in §\ref{groundtruths}). The $x$ axes of these plots correspond to these reference values. To aid visualisation, 10 uniform bins were constructed along each parameter dimension, into which clinical voxels were assigned based on their ``groundtruth" parameter values. Fig \ref{fig:clinicallowsnr} plots the mean bias, standard deviation, and RMSE associated with each bin as a function of the bin's central value, together with the distribution of voxels across the 10 bins.

\begin{figure}[!h]
  \includegraphics[width=\textwidth]{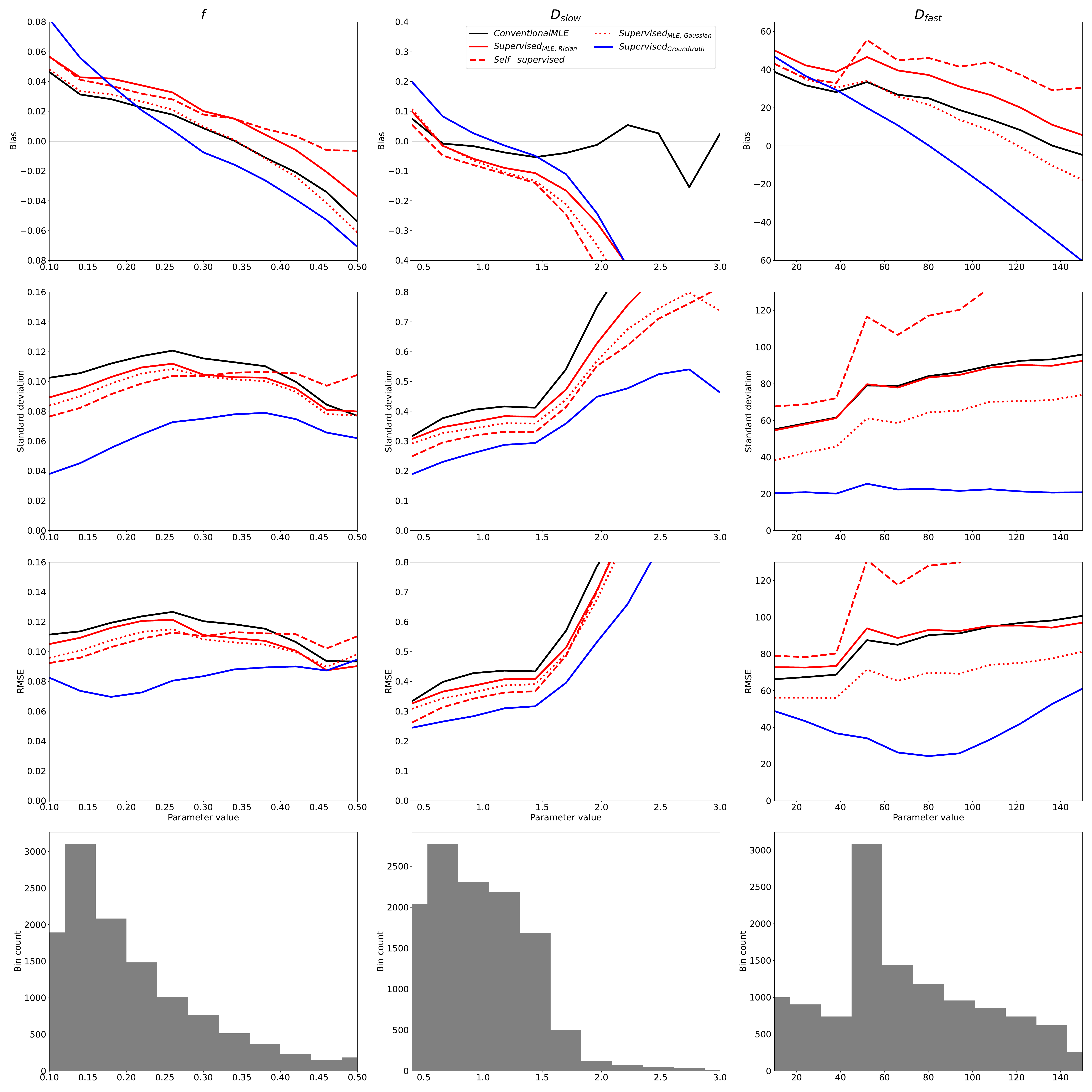}
  \caption{Parameter estimation performance of networks trained on low SNR (15) synthetic data, tested on a synthetic dataset matching the distribution of in vivo reference parameter values. The first three rows summarise performance by showing bias \& RMSE with respect to groundtruth value and standard deviation with respect to noise repetition, marginalised over all non-specified ${Y}$ dimensions. The bottom row shows the distribution of groundtruth parameter values across the parameter range, which matches the in vivo dataset by construction.}
  \label{fig:clinicalsynthlowsnr}
\end{figure}

By calculating the variance of the 16 $b=0$ images, the SNR of this clinical dataset was found to be $\sim$15; Fig \ref{fig:lowsnr} is therefore the relevant point of comparison. It can be readily seen that the trends observed in simulated data, described in §\ref{summary}, are replicated for $f<0.40$, $D_{slow}<1.5$, and the entire range of $D_{fast}$, namely the regions of parameter-space which are well-represented in the real-world data. Fig \ref{fig:clinicalsynthlowsnr} confirms that divergence outside of these ranges is due to under-representation in the in vivo test data; the apparent divergences can be replicated in-silico by matching real-world parameter distributions. 

\begin{figure}[!h]
  \includegraphics[width=\textwidth]{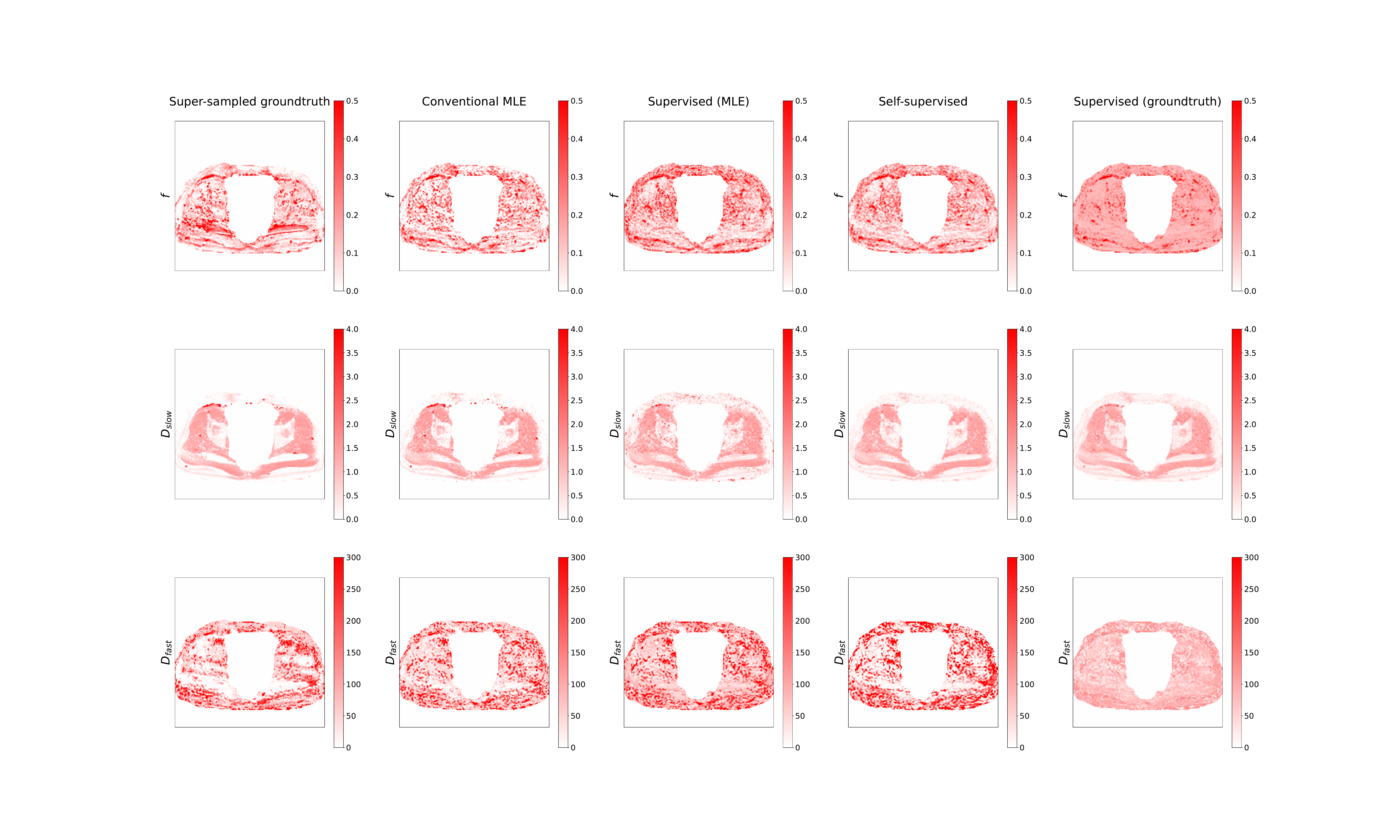}
  \caption{Parameter estimation performance of networks on real-world test data, visualised as spatial maps. Groundtruth maps are taken as the maximum likelihood parameter estimates associated with the complete 160 b-value dataset, whereas network predictions are obtained from a single 10 b-value subsample.}
  \label{fig:clinicalmaps}
\end{figure}

Fig \ref{fig:clinicalmaps} contains exemplar parameter maps from the clinical test data, and shows the real-world implications of the trends summarised in Figures \ref{fig:lowsnr} and \ref{fig:clinicallowsnr}: $Supervised_{GT}$'s low-variance, low-RMSE parameter estimation results in artificially smooth IVIM maps biased towards mean parameter values.

\subsection{Advantages offered by our method}

Our proposed method occupies the low-bias side of the bias-variance trade-off discussed in §\ref{groundtruths}, and offers four broad advantages over the competing method in this space (\textit{Self-supervised}): (i) flexibility in choosing inter-parameter loss weighting $W$, (ii) incorporation of non-Gaussian (e.g. Rician) noise models, (iii) compatibility with complex, non-differentiable signal models $M$, and (iv) ability to interface with low-variance methods, to produce a hybrid approach tunable to the needs of the task at hand. These advantages are analysed in turn.

\subsubsection{Choice of inter-parameter loss weighting $W$}

\begin{figure}[!h]
  \includegraphics[width=\textwidth]{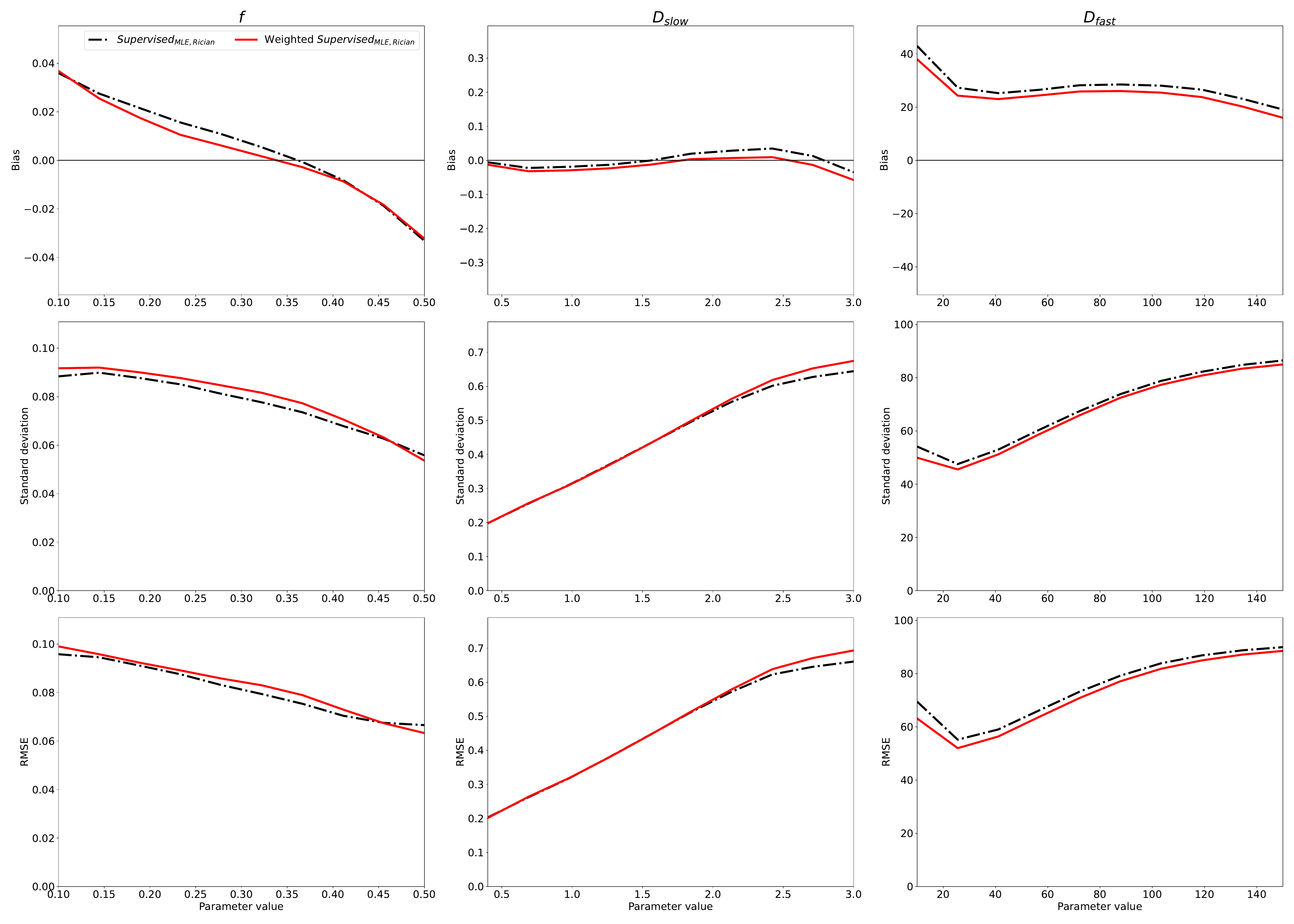}
  \caption{Comparison between \textit{Supervised\textsubscript{MLE, Rician}}, as described above, and variants which differ in their inter-parameter loss weighting $W$, at low SNR (15). Each column compares \textit{Supervised\textsubscript{MLE, Rician}} to a different network variant, uniquely trained to overweight the single relevant signal model parameter. For the sake of visualisation, each plotted point represents marginalisation over all non-specified ${Y}$ dimensions.}

  \label{fig:highsnr_weighting}
\end{figure}

\noindent By computing loss in parameter-space $Y$, our method has total flexibility in adjusting the relative contribution of different $y$ in the training loss function. In contrast, since \textit{Self-supervised} calculates training loss in $X$, the relative weighting depends on the acquisition protocol $z$. Fig \ref{fig:highsnr_weighting} compares our method - weighted so as to not discriminate between different model parameters - with variants designed to overweight single parameters by a factor of $10^6$. The potential advantages offered by this selective weighting are seen in the estimation $D_{fast}$, where this approach leads to a small increase in both precision and accuracy.  This parameter-specific weighting is not accessible within a \textit{Self-supervised} framework.

In light of the differences arising from inter-parameter loss weighting, for subsequent analysis we use \textit{Supervised\textsubscript{MLE, Gaussian}} as a proxy for \textit{Self-supervised}; both methods encode the same regularised MLE fitting, but differ in their inter-parameter weighting.

\subsubsection{Incorporation of Rician noise modelling}

\begin{figure}[!h]
  \includegraphics[width=\textwidth]{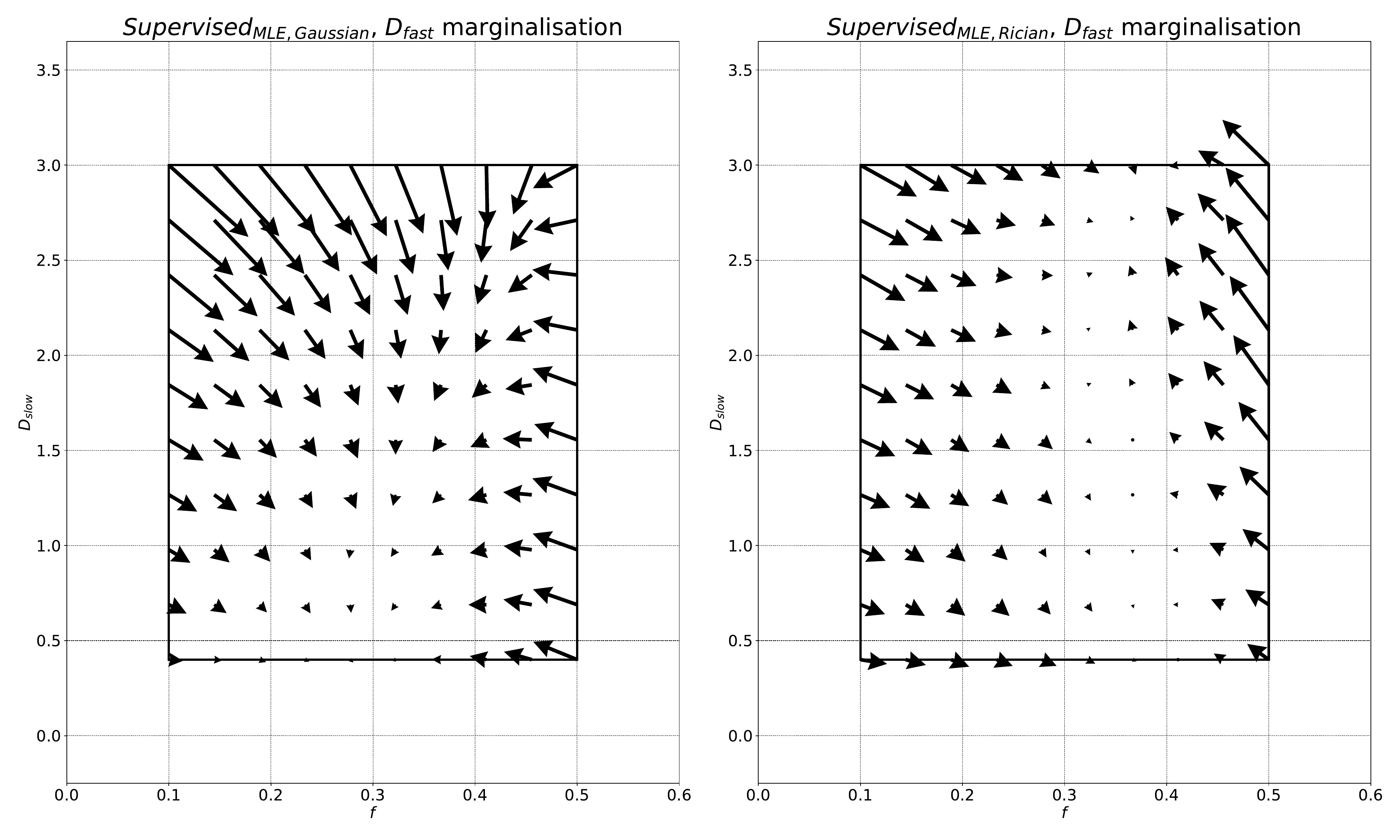}
  \caption{Comparison of the information content captured by \textit{Supervised\textsubscript{MLE} methods}, as a function of the noise model used in computing MLE labels, at low SNR (15). Arrows represent the mean mapping from $Y$ to $\hat{Y}$, averaged over noise, as a function of parameter space $Y$. For the sake of visualisation, each plotted point represents marginalisation over all non-specified $Y$ dimensions.}
  \label{fig:quiver2}
\end{figure}

\noindent By pre-computing MLE labels using conventional parameter estimation methods, we are able to incorporate accurate Rician noise modelling. Comparison between \textit{Supervised\textsubscript{MLE, Rician}} and \textit{Supervised\textsubscript{MLE, Gaussian}} shows the effect of the choice of noise model; these differences are most pronounced at low SNR (Fig \ref{fig:lowsnr}) and high $D_{slow}$, when the Gaussian approximation of Rician noise is known to break down. In this regime, our method gives less biased, more informative $D_{slow}$ estimates, replicating conventional MLE performance at a fraction of the computational cost. At high $D_{slow}$, our method has a flatter, more information-rich, $D_{slow}$ bias curve than all other DL methods. This information loss is further visualised in Fig \ref{fig:quiver2}, which shows the compression in $D_{slow}$ estimates $\hat{X}$ over the groundtruth parameter-space $X$. As expected, this compression is most apparent at high values of $D_{slow}$, when the signal is more likely to approach the Rician noise floor.

\subsubsection{Compatibility with complex signal models}

\noindent An additional advantage of computing training loss in parameter-space $Y$ is that DNN networks are signal model agnostic: network training does not require explicit calculation of $M$. This approach is advantageous when working with complex signal models, as made clear by comparison with \textit{Self-supervised} methods. In contrast with our proposed approach, \textit{Self-supervised} methods embed $M$ between network output and training loss (see Fig \ref{fig:methodcomparison}); this poses two practical limitations. 

The first relates to efficient implementation of mini-batch loss, which requires a vectorised representation (and calculation) of predicted signals. This may pose a non-trivial challenge in the case of complex signal models. The second limitation relates to how training loss is minimised: network parameters $p$ are updated by computing partial derivatives of the training loss. This process requires the loss to be expressed in a differentiable form; embedding $M$ in the loss formulation limits \textit{Self-supervised} methods to signal models that can be expressed in an explicitly differentiable form. 

Our method sidesteps both limitations by not requiring explicit calculation of $M$ during training, and is therefore compatible with a wider range of complex qMRI signal models. 

\subsubsection{Tunable network approach}
\begin{figure}[!h]
  \includegraphics[width=\textwidth]{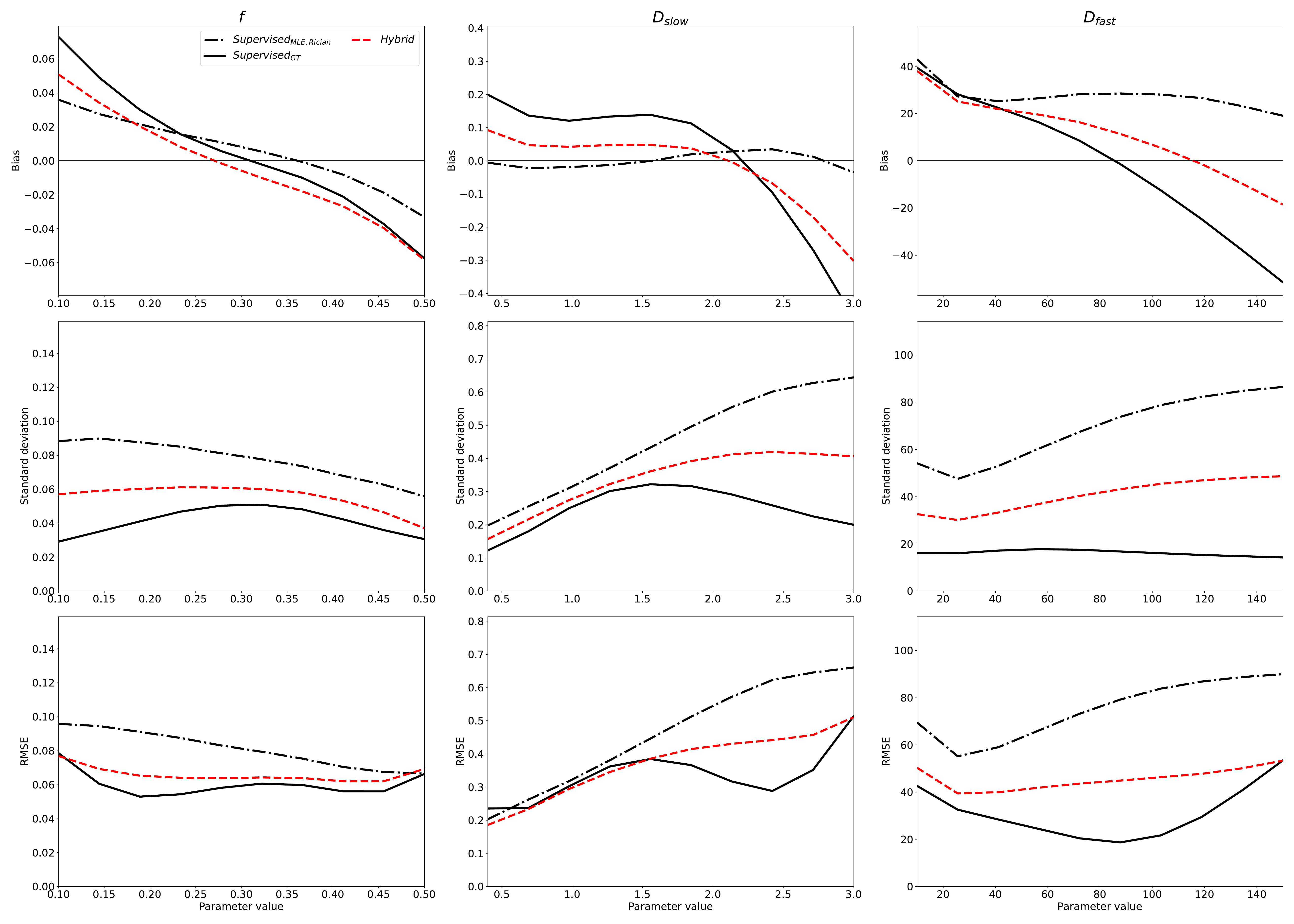}
  \caption{Proof of concept of a hybrid parameter estimation method, formed by training a supervised network with an equally-weighted sum of \textit{Supervised\textsubscript{MLE, Rician}} and \textit{Supervised\textsubscript{GT}} loss functions ($\alpha=0.5$), at low SNR (15). For the sake of visualisation, each plotted point represents marginalisation over all non-specified $Y$ dimensions.}
  \label{fig:hybrid}
\end{figure}

\noindent As discussed above, we show a clear bias/variance trade-off between different parameter estimation methods. The optimal choice of method depends on the task at hand \citep{Epstein2021}, and may not lie at either extreme of this trade-off. Therefore, it would be advantageous to be able to combine low-bias and low-variance methods into a single, hybrid approach, with performance tunable by the relative contribution of each constituent method. Our proposed method, which interfaces naturally with $Supervised_{GT}$, offers exactly that. An example of this approach is shown in Fig \ref{fig:hybrid}: training loss has been weighted equally ($\alpha=0.5$) between groundtruth and MLE labels, and, as expected, the resulting network performance lies in the middle ground between these two extremes.

\subsubsection{Comparison with conventional fitting}

\noindent Comparison between our proposed method (\textit{Supervised\textsubscript{MLE, Rician}}) and conventional fitting (\textit{MLE, Rician}) highlights additional advantages offered by our approach. Firstly, Figs \ref{fig:lowsnr} and \ref{fig:highsnr} demonstrate qualitatively similar performance between these methods across the entire parameter space. The fact that our method, which offers near-instantaneous parameter estimation, produces similar parameter estimates to well-understood conventional MLE methods justifies its adoption in and of itself. However, our method not only mimics but indeed in many cases outperforms (lower bias, variance, and RMSE) the very same method used to compute those labels. This result not only motivates its use, but also confirms that DL methods are able to exploit information shared between training samples beyond what would be possible by considering each sample in isolation.

\subsection{A note on RMSE}

We note that RMSE is a poor summary measure of network performance. RMSE is heavily skewed by outliers, and thus favours methods which give parameter estimates consistently close to mean parameter values. Such estimates, as in the case of $D_{fast}$, may contain very little information (Fig \ref{fig:quiver1}) despite being associated with low RMSE. Accordingly, we strongly recommend that RMSE be discontinued as a single summary metric for parameter estimation performance: it must always be accompanied by bias, variance, and ideally an analysis of information content.

RMSE's limitations as a performance metric during \textit{testing} may also call into question its suitability as a loss metric during \textit{training}. This work, much like the rest of the DL qMRI literature, employs a training loss (MSE, described in Sections \ref{sec:existing} and \ref{sec:proposed}) which is monotonically related to RMSE. Whilst outside the scope of this work, implementing a non-RMSE-derived training loss (such as mean absolute error) may be worth of future investigation.

\subsection{Justification of parameter marginalisation} \label{marginalisation}

The above analysis has been largely based on Figs \ref{fig:lowsnr} and \ref{fig:highsnr}, which show parameter estimation performance marginalised over 3 dimensions of $X$. This choice, made to aid visualisation, was validated against higher dimensional representations of the same data. 

\begin{figure}[!h]
  \includegraphics[width=\textwidth]{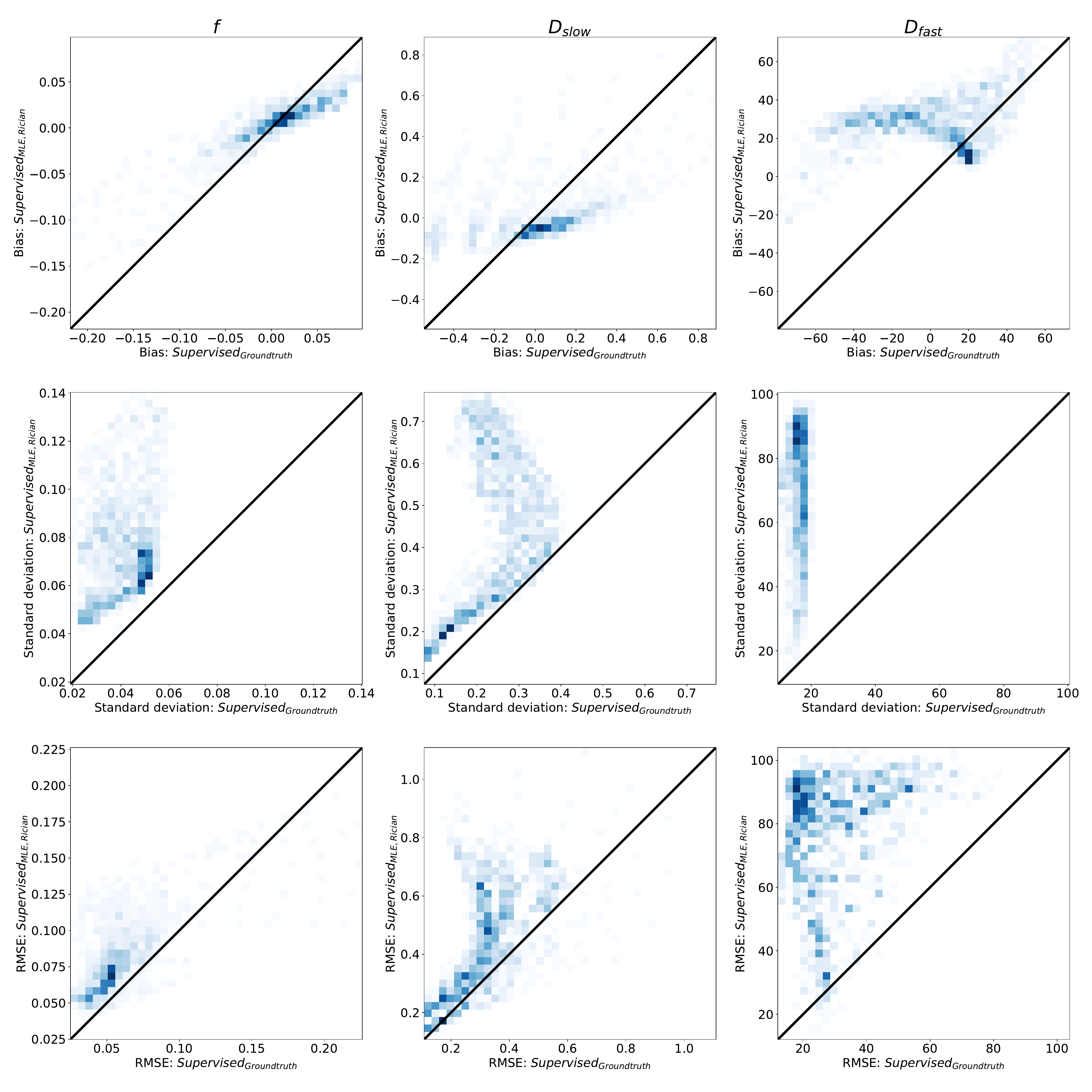}
  \caption{Non-marginalised comparison of parameter estimation performance between \textit{Supervised\textsubscript{MLE, Rician}} and \textit{Supervised\textsubscript{Groundtruth}} at low SNR (15). Colour intensity represents density of distribution across all $X$ and all noise repetitions.}
  \label{fig:noisemodel_scatter}
\end{figure}

\begin{figure}[!h]
  \includegraphics[width=\textwidth]{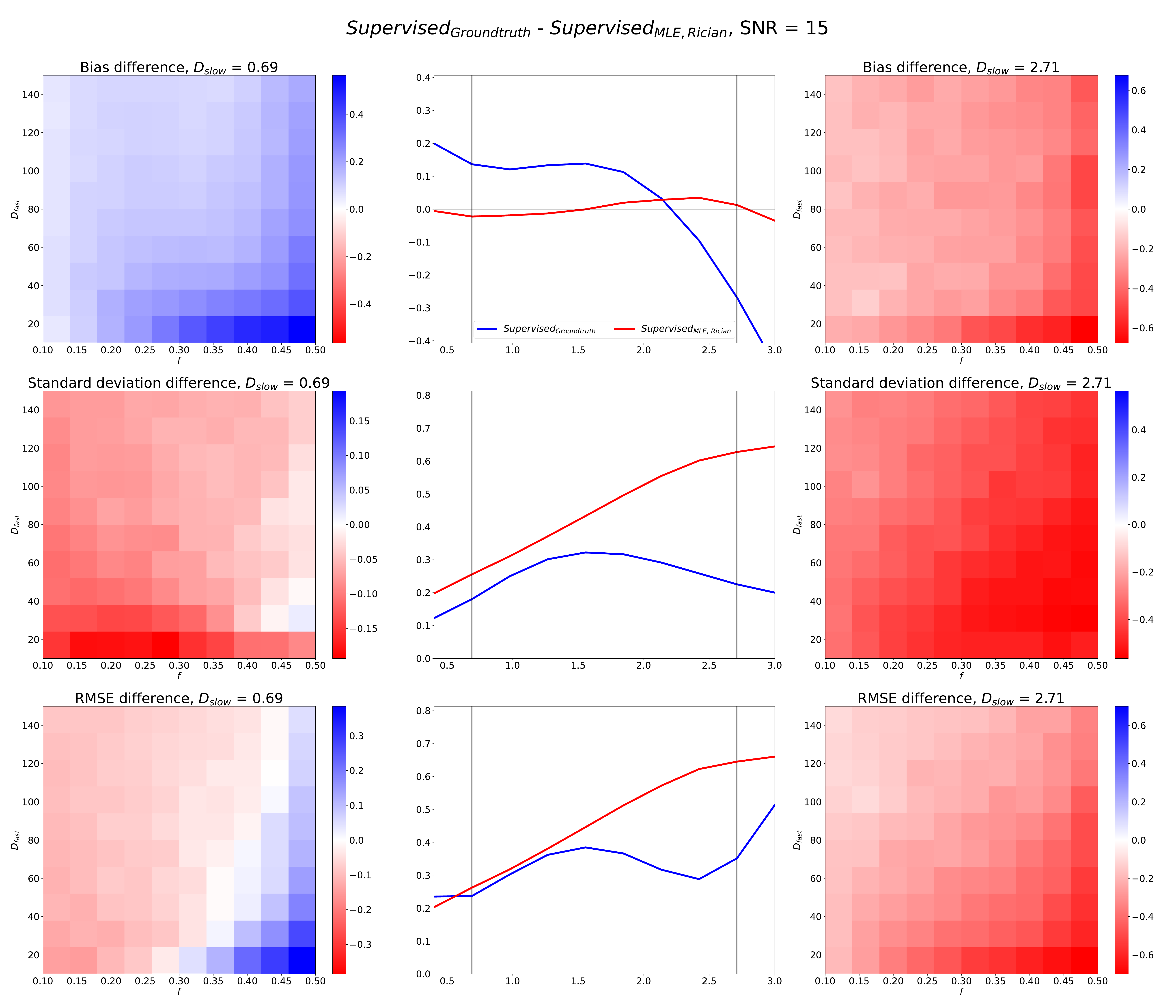}
  \caption{Differences in performance (bias, standard deviation, RMSE) between \textit{Supervised\textsubscript{MLE, Rician}} and \textit{Supervised\textsubscript{Groundtruth}} for two groundtruth values of $D_{slow}$ at low SNR (15). The outermost columns (left and right) correspond to $D_{slow}=0.69$ and $D_{slow}=2.71$ respectively, and show mean performance under noise repetition, without marginalisation. The central column reproduces the corresponding marginalised representation from Fig \ref{fig:lowsnr}.}
  \label{fig:lowsnr_compare_1}
\end{figure}

Fig \ref{fig:noisemodel_scatter} compares  \textit{Supervised\textsubscript{MLE, Rician}} and \textit{Supervised\textsubscript{Groundtruth}} performance across the entire qMRI parameter space. It can be seen that trends observed in Fig \ref{fig:lowsnr} are replicated here; we draw attention to two such examples. Firstly, Fig \ref{fig:lowsnr} suggests \textit{Supervised\textsubscript{Groundtruth}} produces lower $f$ standard deviation than \textit{Supervised\textsubscript{MLE, Rician}}; Fig \ref{fig:noisemodel_scatter} confirms this to be the case across all test data. In contrast, Fig \ref{fig:lowsnr} suggests that \textit{Supervised\textsubscript{Groundtruth}} produces higher $D_{slow}$ bias at low $D_{slow}$ and lower bias at high $D_{slow}$; Fig \ref{fig:noisemodel_scatter} confirms a spread of bias differences across the test data: some favouring one method, and others the other. This effect is explored in Fig \ref{fig:lowsnr_compare_1}, which compares $D_{slow}$ estimation performance as a function of $f$ and $D_{fast}$ at two specific (non-marginalised) groundtruth $D_{slow}$ values ($0.69,2.71)$. As expected from the marginalised representation in Fig \ref{fig:lowsnr}, at low $D_{slow}$ \textit{Supervised\textsubscript{Groundtruth}} produces higher bias \textit{across the entire} $f$-$D_{fast}$ \textit{parameter space}, whereas at high $D_{slow}$ the opposite is true.

Despite this, it is important to note the limitations of marginalisation. Fig \ref{fig:lowsnr_compare_1} also shows that the relative performance of \textit{Supervised\textsubscript{MLE, Rician}} and \textit{Supervised\textsubscript{Groundtruth}} varies across \textit{all parameter-space dimensions}. Consider $D_{slow}=0.69$, where \ref{fig:lowsnr} shows similar marginalised RMSE for these methods. In fact, by visualising this difference as a function of $f$ and $D_{fast}$, we reveal two distinct regions: high $f$/low $D_{fast}$ (where \textit{Supervised\textsubscript{MLE, Rician}} produces lower RMSE), and elsewhere (where it produces higher RMSE). This highlights (i) the potential pitfalls of producing summary results by marginalising across entire parameter spaces and (ii) the need to choose parameter-estimation methods appropriate for the specific parameter combinations relevant to the tissues being investigated \citep{Epstein2021}.

\subsection{Non-voxelwise approaches}

This work has focused on voxelwise DL parameter estimation methods: networks which map one signal curve to its corresponding parameter estimate. There are, however, alternatives: convolutional neural network methods which map spatially related clusters (``patches") of qMRI signals to corresponding clusters of parameter estimates \citep{Fang2017, Ulas2019, Li2022}. Our MLE training label approach could be incorporated into such methods, and we leave it to future work to investigate the effect this would have on parameter estimation performance.

\section{Conclusions}
\label{sec:conclusions}

In this work we draw inspiration from state-of-the-art supervised and self-supervised qMRI parameter estimation methods to propose a novel DNN approach which combines their respective strengths. In keeping with previous work, we demonstrate the presence of a bias/variance trade-off between existing methods; supervised training produces low variance under noise, whereas self-supervised leads to low bias with respect to groundtruth. 

The increased bias of supervised DNNs is counter-intuitive - when labels are available, these methods have access to more information, and should therefore outperform, non-labelled alternatives. In light of this, we infer that the high bias associated with these supervised methods stems from the \textit{nature} of the additional information they receive: groundtruth training labels. By substituting these labels with independently-computed maximum likelihood estimates, we show that the low-bias performance previously limited to self-supervised approaches can be achieved within a supervised learning framework. 

This framework forms the basis of a novel low-bias supervised learning approach to qMRI parameter estimation: training on conventionally-derived maximum likelihood parameter estimates. This method offers four clear advantages to competing non-supervised low-bias DNN approaches: (i) flexibility in choosing inter-parameter loss weighting, which enables network performance to be boosted for qMRI parameters of interest; (ii) incorporation of Rician noise modelling, which improves parameter estimation at low SNR; (iii) separation between signal model and training loss, which enables the estimation of non-differentiable qMRI signal models;and, crucially, (iv) ability to interface with existing supervised low-variance approaches, to produce a tunable hybrid parameter estimation method. 

This final point underpins the key contribution of this work: unifying low-bias and low-variance parameter estimation under a single supervised learning umbrella. When faced with a parameter estimation problem, we no longer need to choose between extremes of the bias/variance trade-off; we can now tune DNN parameter estimation performance to the specific needs of the task at hand. This sets the stage for future work, where this tuning constant is optimised as part of a computational, task-driven, experimental design framework \citep{Epstein2021}.

\acks{SCE is supported by the EPSRC-funded UCL Centre for Doctoral Training in Medical Imaging (EP/L016478/1). TJPB is supported by an NIHR Clinical Lectureship (CL- 2019-18-001) and, together with MHC, is supported by the National Institute for Health Research (NIHR) Biomedical Research Centre (BRC). This work was undertaken at UCLH/UCL, which receives funding from the UK Department of Health’s the NIHR BRC funding scheme.}

%
\ethics{The work follows appropriate ethical standards in conducting research and writing the manuscript, following all applicable laws and regulations regarding treatment of animals or human subjects.}

\coi{The authors confirm they have no conflict of interest to disclose.}

\data{Data and code are available at~\url{https://github.com/seancepstein/training_labels}.}

\bibliography{bibliography}





\end{document}